\documentclass[aps,prb,reprint,amssymb,amsfonts,longbibliography]{revtex4-2}

\pdfoutput=1

\usepackage[english]{babel}
\usepackage{amsmath}
\usepackage{amssymb}
\usepackage{bm}
\usepackage{graphicx}
\usepackage[unicode]{hyperref}
\usepackage[latin1]{inputenc}
\usepackage{array}
\usepackage{upgreek}
\usepackage{color}
\usepackage{todonotes}

\begin{document}

\title{Increased muon field at surface and substrate interface of palladium thin films}
	%
\author{G. Welker$^1$} 
\email{corresponding author: g.welker@gmx.net}
\author{M. de Wit$^1$} 
\author{T. H. Oosterkamp$^1$}
\author{J. A. Mydosh$^1$}
\author{A. Suter$^2$}
\author{T. Prokscha$^2$}
\author{L. Bossoni$^{3}$}

\affiliation{1 Leiden Institute of Physics, Leiden University, PO Box 9504, 2300 RA Leiden, The Netherlands}
\affiliation{2 Laboratory  for  Muon  Spin  Spectroscopy,  Paul  Scherrer  Institute,  CH-5232  Villingen  PSI,  Switzerland}
\affiliation{3 C.J. Gorter Center for High Field MRI, Department of Radiology, Leiden University Medical Center, Albinusdreef 2, 2333 ZA Leiden, The Netherlands}

\date{\today}

\begin{abstract}
We performed depth-dependent low-energy muon spin spectroscopy ($\mu$SR) studies on three palladium 100 nm thin films, both undoped and doped with 170 ppm of iron. Muons implanted in the surface and substrate interface region probe an increased local magnetic field compared to the inner part of the sample. The field increase extends over a few nanometers, it is temperature-independent (in the range of 3.7 - 100 K), stronger for the iron-doped samples and accompanied by an increase in local field inhomogeneity. We consider various potential origins for this magnetic surface state, such as adsorbents and supressed d-states. Our conclusion is that orbital moments induced at the surface / interface by localized spins and charges are the most likely explanation, potentially accompanied by magnetic moments due to crystal irregularities.

\end{abstract}

\maketitle

\section{Introduction}
Elemental palladium (Pd) is a versatile and intriguing transition metal that exhibits a variety of interesting and practical properties both as pure or when combined with other elements. For example, Pd is an incipient ferromagnet metal that shows magnetic spin glass behavior and giant magnetic moments when doped with small amounts of magnetic elements such as iron (Fe) \cite{Mydosh1980}. Such magnetically doped Pd can be used for millikelvin thermometry \cite{Jutzler1986}. Pd nanomaterial has also long been known for its abilities to purify, store and detect hydrogen \cite{Adams2011}. It plays a major role as a catalyst for different applications by exploring e.g. metal organic frameworks \cite{Yang2019}. In multilayer thin film structures, Pd is used in magnetoresistance spin-valves \cite{Gouda1999} and in phase coherence superconducting junctions \cite{Glick2017}.\\
Most recently, Pd and other transition metals have become of interest in the emerging field of spin-orbitronics \cite{Manchon2017}, where spin-orbit coupling drives a variety of magnetic phenomena at surfaces and interfaces. Pd surfaces and interfaces can exhibit vastly different physical behavior than bulk Pd. This is of great interest considering the above-mentioned large number of applications for Pd nanomaterial and high surface to volume ratios for nanoscale objects. Pd nanoparticles can for example be either paramagnetic with varying susceptibilities \cite{VanLeeuwen1992} or even ferromagnetic \cite*{Taniyama1997,Shinohara2003,Sampedro2003}, while bulk Pd is a Stoner-enhanced paramagnet. Several experimental studies demonstrated a size dependence for Pd nanoparticle magnetic properties and/or emphasized the importance of the nanoparticle's surface \cite*{Ladas1978,Wertheim1986,VanLeeuwen1992,Taniyama1997,Shinohara2003,Hernando2006,Crespo2013,Asensio2017,Hooper2018}. Furthermore, Beta-NMR measurements on Pd thin films revealed the existence of a magnetic surface state with a positive lithium Knight shift \cite{MacFarlane2013}.\\
These experimental studies come to different conclusions about the origin of their measured magnetic Pd surface state. Some of them cannot unambiguously identify a single underlying mechanism. Explanations include changes in the density of states due to the reduced coordination number at the surface \cite*{Shinohara2003} or adsorbed molecules \cite{Crespo2013}, crystal defects \cite*{Alexandre2006,Kulriya2012} and orbital magnetic moments \cite{Hernando2006}. Reported experimental length scales for these surface states vary between a few atomic layers \cite{Shinohara2003} and more than 22.5 nm \cite{MacFarlane2013}. Accordingly, a systematic study of a magnetic Pd surface state is needed to clarify the role and interplay of potential physical mechanisms.\\
We have fabricated a series of sputtered Pd films for use in muon implantation to study internal magnetic fields and depolarization rates of $\mu$SR response at various muon implantation depths and temperatures.  Pd and Pd(Fe) films with a nominal thickness of 100 nm were prepared on silicon (Si) substrates, one sample was prepared with an interstitial gold (Au) layer at the Si interface. Muon spin rotation spectroscopy is our basic tool to study the magnetic properties of the Pd films. Muons with variable implantation energies were employed to probe the films at different depths in order to examine the surface behavior, the bulk-like properties in the center of the 100 nm film and the Pd-(Au)Si interface. The muon depolarization rate and the local magnetic field at the muon stopping site showed variations in different regions of the thin films. While the bulk-like areas of the film exhibited properties comparable to previous muon experiments on Pd(Fe) \cite*{Nagamine1977,Nagamine1977a}, both surface and interface regions were characterized by an increased muon field. We ascribe this to the existence of a magnetic surface state likely caused by orbital magnetic moments. 

\section{Methods}
\subsection{Sample production and morphology}
\begin{figure}
\centering
\includegraphics[width=\columnwidth]{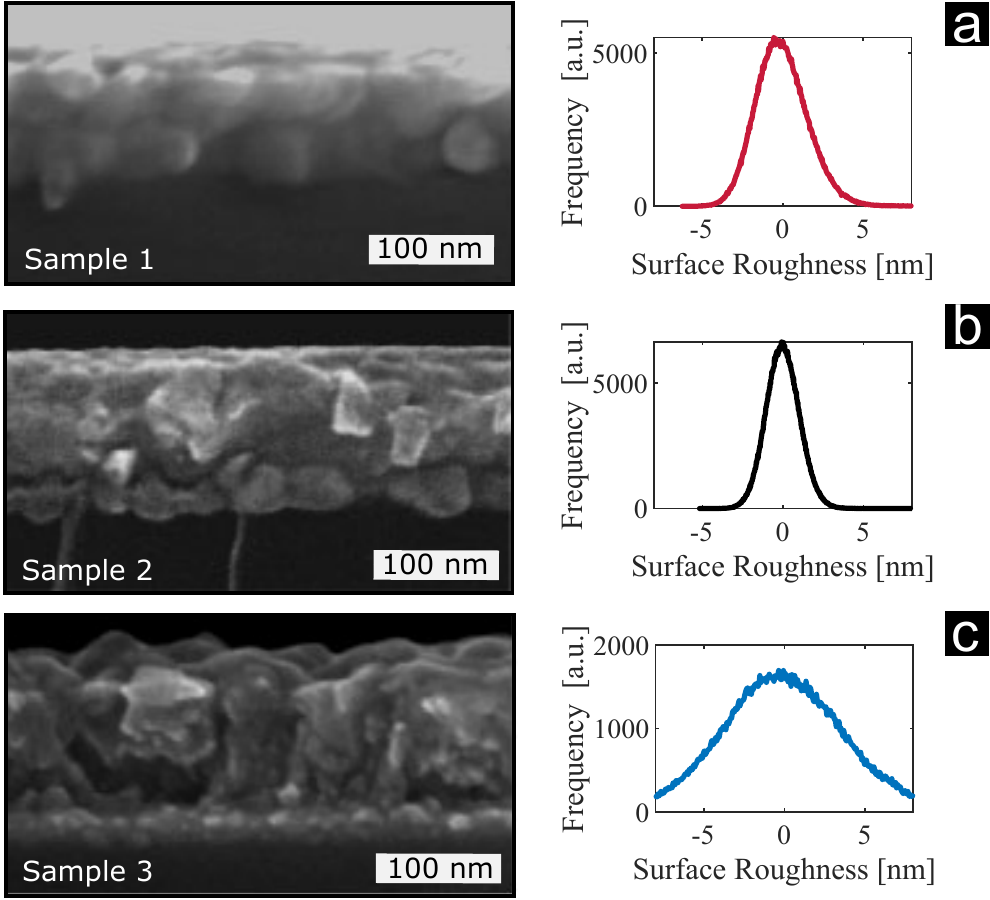}
\caption{Overview of sample properties. SEM images of the cleaved samples (side view, Si substrate visible at the bottom) and surface roughness profiles derived from AFM measurements for (a) Pd, (b) Pd(Fe) and (c) Pd(Fe)Au.}
\label{figure:Sample_picture}
\end{figure}

Three Pd thin films were grown on an intrinsic silicon substrate by means of radio frequency (RF) sputtering. An overview of all samples is shown in Figure (\ref{figure:Sample_picture}). The diced substrate pieces were first dipped for 4-5 seconds into hydrofluoric (HF) acid to remove oxides, followed by three H$_2$O baths and dry-blowing with nitrogen. Then the silicon was loaded into a UHV system as quickly as possible (less than 20 minutes) to reduce the formation of new oxides. We started the sputtering at a chamber pressure of $7.7 \cdot 10^{-9}$ mbar, using an argon pressure of $3.3 \cdot 10^{-3}$ mbar and an RF power of $40.1$ W ($100$ mA, $401$ V). The Pd target for sample 1 with 99.99\% purity purchased from ESPI is specified to contain less than 2 ppm of iron and no other magnetic impurities. The Pd sputter target for sample 2 and 3 was custom-made by Goodfellow, starting from Pd with 99.99\% purity and magnetic impurity levels of cobalt, chromium and nickel confirmed to be below 10 ppm. The final target was a Pd-iron alloy with an iron doping of 170 ppm. The iron content of the alloy target was confirmed by inductively coupled plasma mass spectrometry (ICP analysis). \\
As we are interested in the magnetic properties of the Pd-substrate interface, sample 3 has a gold spacing layer between the Pd and the silicon substrate. This was done by evaporating 20 nm of gold on the silicon after the HF dip. The sample was then exposed to air while placing it into the UHV system for the sputtering of the Pd.\\
We measured the thickness of the Pd films with a DektakXT profilometer. The pure Pd film in sample 1 has an average thickness of 96 nm, the iron-doped Pd films with and without gold have an average  thickness of 105 nm. Imaging the films in a scanning electron microscope revealed a grainy structure with grain diameters of tens of nanometers. The grain diameter median is 40 nm for the Pd(Fe) sample and 30 nm for the Pd(Fe)Au sample. The samples were also characterized with atomic force microscopy, the surface roughness profiles are shown in Figure (\ref{figure:Sample_picture}).\\

\subsection{Magnetic impurity characterization:\\ SQUID magnetometry}
\begin{figure}[t]
\centering
\includegraphics[width=\columnwidth]{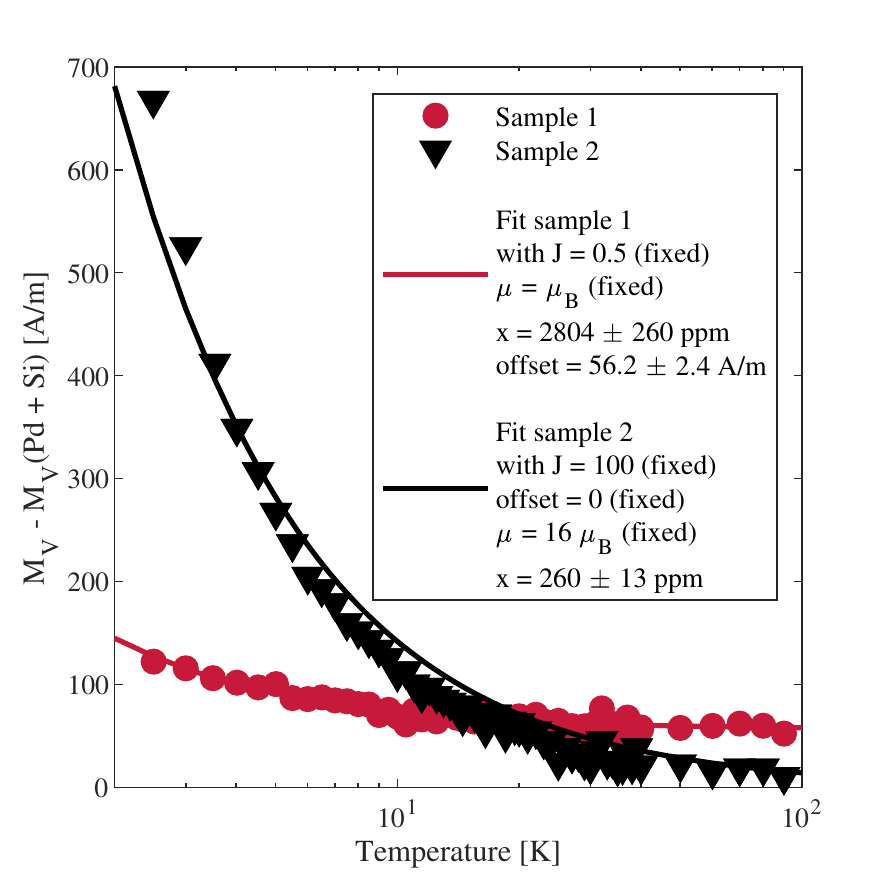}
\caption{Impurity contribution to sample magnetization in an applied field of 150 mT. Equation (\ref{Brillouin}) is used as a fitting function to extract the impurity concentration $x$ from the SQUID magnetometry data. Sample 1 is dominated by localized single electron spins, sample 2 is dominated by iron impurities.}
\label{figure:SQUID}
\end{figure} 

Sample 1 (pure Pd) and sample 2 (iron-doped Pd without gold) were characterized by means of superconducting quantum interference device (SQUID) magnetometry. SQUID measurements can be used to determine the total magnetization, by which one can estimate the concentration of magnetic impurities.\\
We performed SQUID magnetometry measurements using a standard Quantum Design MPMS-XL in reciprocating sample option (RSO) mode. Before the measurements, we manually cleaved the samples into pieces of about 3.2x3.2 mm$^2$ followed by sonicating them in isopropanol to remove potential contaminations and debris from the breaking process. We took great care to avoid magnetic contamination, for example by using a ceramic knife and plastic tweezers. The cleaved samples were mounted in a clear drinking straw and kept in place with folded drinking straws symmetrically placed on both sides. In order to increase the signal-to-noise ratio, we stacked  multiple sample pieces in one straw.\\
We cooled each sample to 2 K, then applied an external magnetic field of 150 mT and measured the magnetization while warming up to 300 K. This measured magnetization contains contributions from the Si, the Pd and the magnetic impurities. A complicating factor in fitting the raw data is that the paramagnetic contribution from the Pd film and the diamagnetic contribution from the Si substrate partially cancel each other. We therefore determined the magnetization of the Si substrate in a separate measurement and subtracted the Si raw data from the (Pd + Si + impurities ) raw data for each temperature. We then extracted a value for the temperature-dependent (Pd + impurities) magnetization by performing a fit to the subtracted data following the procedure outlined in reference \cite{MPMS2002}.\\
For singling out the impurity contribution, we made use of the fact that this contribution only becomes relevant at low temperatures \cite{Herrmannsdorfer1996}. At high temperatures, the measured magnetization stems exclusively from the Pd. Therefore, we subtracted a high temperature average value (measured at 250 - 300K) from the (Pd + impurities) data. The two resulting data sets are displayed in Figure (\ref{figure:SQUID}). They show the temperature dependence of the impurity magnetization and
they resemble the results from a similar measurement conducted by Herrmannsdoerfer et al. \cite{Herrmannsdorfer1996}.\\ 
For both samples, the magnetization can be described by a Brillouin function, which formulates the behavior of non- (or weakly) interacting magnetic moments:
\begin{align}
\begin{split}
M_V(J,\xi) & = M_{sat} \cdot \left\{ \left(\frac{2J+1}{2J}\right) \cdot \coth \left(\frac{\xi \cdot (2J+1)}{2J}\right) \right.\\
						& \left. \qquad - \frac{1}{2J} \cdot \coth \left(\frac{\xi}{2J}\right)  \right\}, \label{Brillouin} \\
\end{split}
\\
& \hspace{18mm} \mathrm{with} ~  M_{sat} = \frac{N_0 \cdot x \cdot \mu}{V_{mol}} \tag{1a}\\
& \hspace{18mm} \mathrm{and} ~ \xi = \frac{\mu B}{k_B T} \tag{1b}.
\end{align}
$M_V$ is the volume magnetization, $N_0$ Avogadro's constant, $x$ the impurity concentration, $\mu$ the magnetic moment of the impurity, $V_{mol}$ the molar volume, $J$ the total angular momentum quantum number, $\mu_B$ the Bohr magneton, $B$ the applied magnetic field, $k_B$ the Boltzmann constant and $T$ the temperature of the sample \cite{Herrmannsdorfer1996}. We determine the impurity concentration in sample 1 and 2 by using Equation (\ref{Brillouin}) as a fitting function with the fitting parameters $x, \mu, J$ and an added offset.\\
Based on sputter target specifications and sample handling in ambient conditions, we expect two major magnetic impurity contributions: Iron doping atoms that polarize palladium d-holes and thereby introduce so-called giant magnetic moments (GMM, $\mu = 13 \mu_B -16 \mu_B , J \geq 100$) \cite{Herrmannsdorfer1996} and localized single electron spins originating from surface adsorbents ($\mu = \mu_B , J = 0.5$).\\
For sample 1, various starting and boundary values for $x, \mu$, $J$ and offset lead to a fit result of $J \approx 0.5$. This indicates that the signal is dominated by single electron spins and that giant magnetic moments play a negligible role. We therefore fix $\mu = \mu_B$ and $J = 0.5$ and repeat the fit, leading to an electron spin concentration of $2804 \pm 260$ ppm and an offset of $56.2 \pm 2.4$ A/m. Due to the surface adsorbents, many of these electrons are located in a thin layer at the sample surface, which allows converting the 3D-spin concentration $x$ in a 96nm thick sample to a 2D-spin density $\sigma = 4.58 \pm 0.42$ spins/nm$^2$.\\
In sample 2 with 170 ppm of iron, we expect GMM to be the dominant contribution. This is confirmed by the magnitude of the measured magnetization. Unrealistic concentrations of single electron spins would be needed to explain the signal. As the size of $\mu$ for GMM slightly depends on the iron concentration, we fix $J = 100$ and $\mu = 15.5 \mu_B$, the values reported by Herrmannsdoerfer et al. for an iron concentration of 190 ppm. We also fix the offset to zero based on visual data inspection. The resulting iron concentration from the fit is $x=277 \pm 14$ ppm. Due to the above-mentioned concentration dependence of $\mu$ and the fact that changes in $\mu$ and $x$ can partially compensate each other in Equation (\ref{Brillouin}), another scenario is fixing $\mu$ to the biggest possible value of $16 \mu_B$, which leads to a fit result of $ x = 260 \pm 13$ ppm in accordance with values from Herrmannsdoerfer et al.\\
In conclusion, we can state that the SQUID data for sample 1 is dominated by localized single electron spins, for sample 2 it is dominated by iron impurities. The SQUID data for sample 2 suggests a slightly higher magnetic impurity concentration than the iron concentration of the sputter target. As other magnetic impurities in the sputter target are below 10 ppm, this higher impurity concentration probably stems from localized single electrons, which are also expected to be present in sample 2 due to the same sample handling as for sample 1.

\subsection{Low Energy Muon Spin Rotation} \label{chap:LEM-MuSR}
\begin{figure}[t]
\centering
\includegraphics[width=\columnwidth]{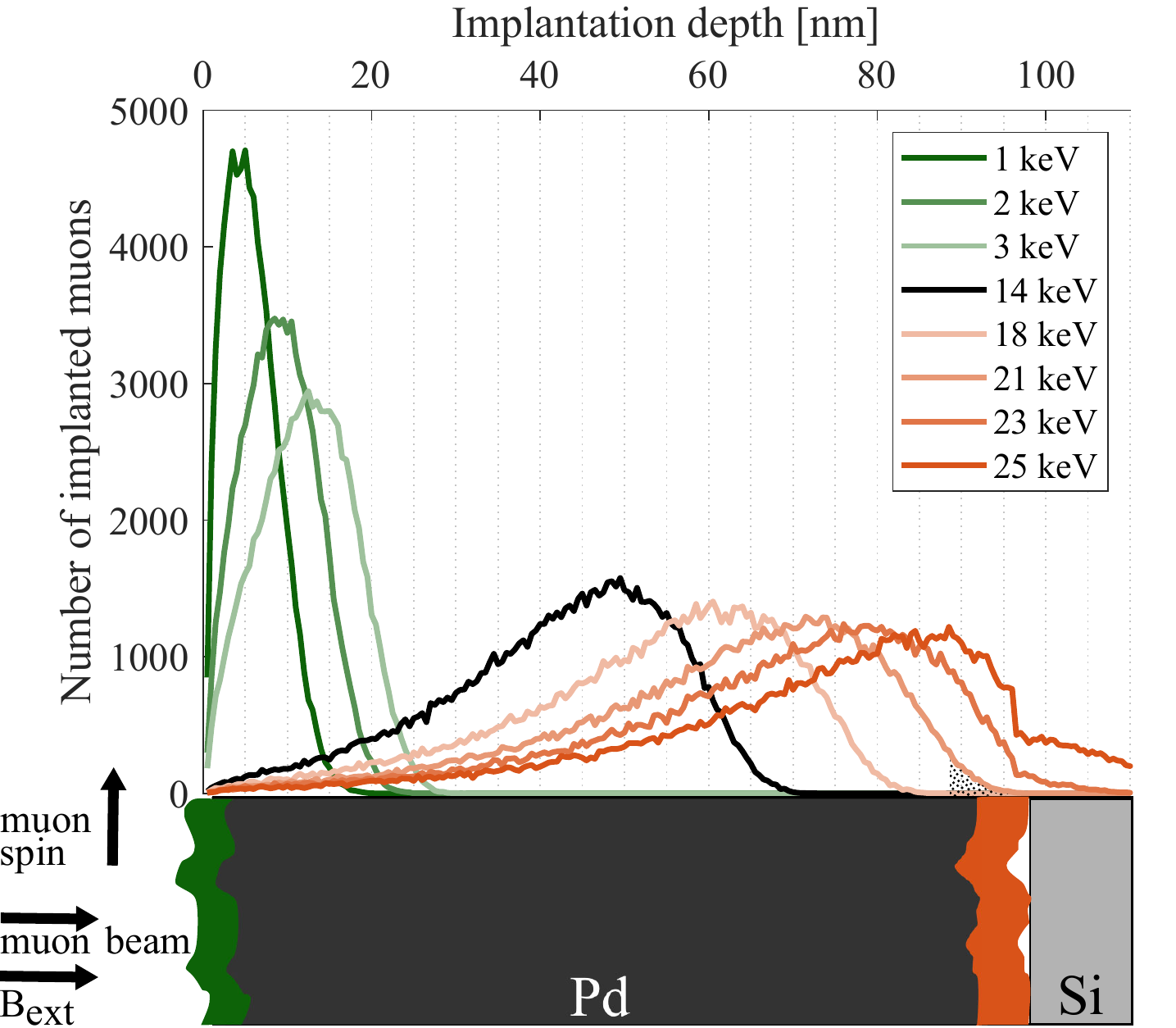}
\caption{Selected implantation profiles simulated with TrimSP \cite*{Morenzoni2002,Eckstein1991} for a 96nm Pd film on a silicon substrate and depiction of the sample structure. For the black implantation profile, no muons are implanted at the surface and interface. For muon energies shown in green (orange), we measured a magnetic field increase in the experiment; the darker the color, the higher the fraction of muons implanting in the surface (interface) region. Notice the sample roughness and the surface (interface) region colored in green (orange). The small dotted area under the stopping profile indicates the fraction of muons that sense interface effects at 21 keV muon energy.}
\label{figure:stopping}
\end{figure}
\subsubsection{Working Principle}
Muon spin rotation experiments probe local magnetic fields inside a sample. In a transversal field $\mu$SR (TF-$\mu$SR) experiment, a spin-polarized beam of positive muons is directed at the sample while an external magnetic field is applied perpendicular to the direction of the spin polarization. The higher the energy of the muons, the deeper they will penetrate into the sample before stopping in the lattice. At the implantation site, the muons precess with the Larmor frequency $\omega_L$ given by:
\begin{align} 
\omega_{L} &= \gamma_{\mu} B_{\mu}, \mathrm{ where}\\
B_{\mu} &= B_{ext} + B_{Lor} + B_{dem} + B_{int}, \label{eq:Bmu}
\end{align}
and the local muon field $B_{\mu}$ at the muon site consists of the external transversal magnetic field $B_{ext}$, the Lorentz field $B_{Lor}$, the demagnetization field $B_{dem}$ and $B_{int}$, the internal field felt by the muon, caused by e.g. the contact hyperfine interaction and local dipolar fields. $\gamma_{\mu} = 2 \pi \cdot 135.5$ MHz/T is the muon gyromagnetic ratio.\\
We are interested in $B_{int}$ as it stems from the direct magnetic environment of the muon inside the sample. It is common to express it in terms of the muon Knight shift, which describes how much the internal field shifts the muon Larmor frequency compared to the situation with only the external magnetic field present. The muon Knight shift is defined as
\begin{equation}
K_{\mu} = \frac{\boldsymbol{B_{ext}} \cdot \boldsymbol{B_{int}}}{B_{ext}^2}.
\end{equation}
We will elaborate on the different muon Knight shift contributions for Pd in the discussion section.\\
The $\mu$SR technique uses the fact that muons are unstable particles with a lifetime of 2.2 $\mu$s, and that the decay positrons are preferentially emitted in the direction of the muon spin. This allows to monitor the muon spin precession and thereby to measure the local muon field. The precessing spin-polarized muon ensemble depolarizes depending on the distribution and / or fluctuations of the local magnetic field in the sample, which provides a way to measure magnetic field inhomogeneities. For samples with dilute impurity concentrations, such as ours, the magnetic field distribution has a Lorentzian shape \cite{Walstedt1974}, leading to an exponential decay of the muon polarization. The time-dependent anisotropic emission of the decay positrons can be written as
\begin{equation}
A(t) = A_0 e^{- \lambda} \cos(\omega_{L} t + \varphi),
\label{eq:poscountrate}
\end{equation}
where $A_0$ is the initial asymmetry (determined by the geometry of the positron detectors), $\lambda$ is the muon spin depolarization rate and $\varphi$ is the phase of the spin precession in the positron detector.
\subsubsection{Measurements performed}
We conducted TF-$\mu$SR measurements on all three samples at the low energy muon facility (LEM) at the $\mu$E4 beamline \cite*{Prokscha2008, Morenzoni2000} at Paul Scherrer Institute. At LEM, polarized positive muons are available with tuneable implantation energies in the range from 1 keV to 30 keV. The implantation energy is mainly adjusted by applying a negative or positive bias voltage to the sample.
In order to determine appropriate muon energies for studying surface- and interface-related effects, we computed depth-dependent muon implantation profiles with the software TrimSP \cite*{Morenzoni2002,Eckstein1991}. As a result, we measured at muon energies between 1 keV and 25 keV at a base temperature of 3.7 K and a transversal magnetic field of 150 mT for all samples. We varied the temperature for all samples from 3.7 K to 100 K at a fixed muon energy of 14 keV. Furthermore, we performed more detailed temperature-dependent measurements on sample 2 at specific muon energies: at 2 keV to study surface effects, at 23 keV to examine the sample-substrate interface and at 14 keV to have a reference measurement without any surface or interface effects present. These implantation energies were chosen based on the TrimSP simulations visible in Figure (\ref{figure:stopping}). It is important to note that even for the lowest (highest) implantation energies, unavoidably only a small fraction of the muons is implanted at the surface (interface), as can also be seen in Figure (\ref{figure:stopping}).\\
We analyzed the raw data with the web-based application \cite{musrfitwebpage} of the musrfit software \cite{Suter2012}, performing a fit to Equation (\ref{eq:poscountrate}). An example of raw data and fit is presented in the inset of Figure (\ref{fig:all_samples_depth} a).

\begin{figure}[t!]
\centering
\includegraphics[height=0.96\textheight]{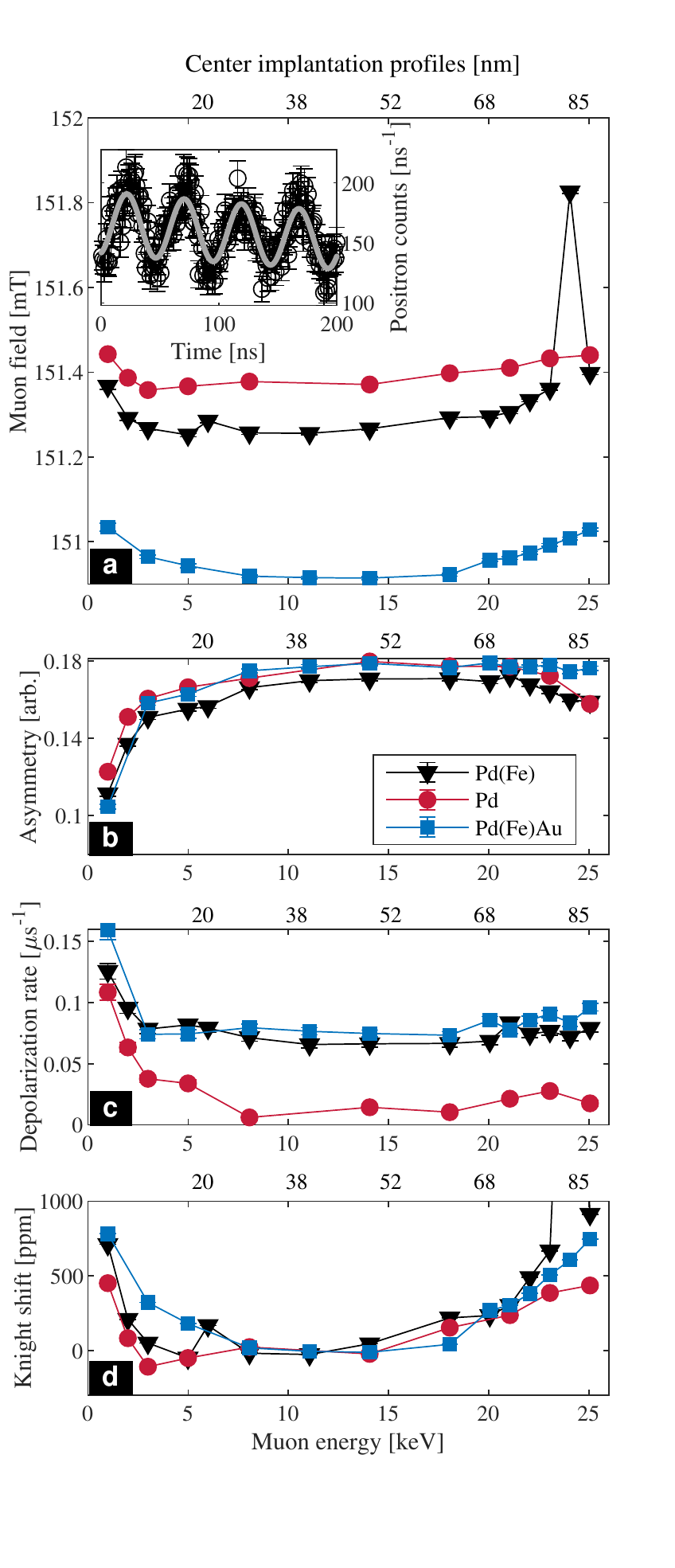}
\end{figure}

\section{Results}

\begin{figure}[b!]
\par\noindent\rule{\columnwidth}{0.4pt}
 \caption{(Figure to the left): Muon field (a), asymmetry (b), depolarization rate (c) and Knight shift (d) versus implantation energy for all samples at 3.7K. The inset shows raw data and fit for the 14 keV data point of the Pd(Fe) sample. The Pd(Fe)Au sample was measured in a different cryostat, therefore muon field values are lower than for the other two samples. The Pd(Fe) data has two outliers with an increased muon field, 6 keV and 24 keV. These data points were measured after other measurements at 330 mT, the remanence of the experiment magnet caused the increased muon field. In panel (d), the Knight shift is calculated with the field values in the sample center (averaging 8-14 keV) as a reference point. Error bars depict the fitting error and are mostly smaller than the marker size.}
\label{fig:all_samples_depth}
\end{figure}

In our TF-$\mu$SR measurements, we found that both the measured muon field and the depolarization rate vary with implantation depth and sample temperature. The most striking result is an increase in muon field and depolarization rate at the surface and interface of the samples. In this section, we primarily describe sample-induced effects, we consider the depth dependence and temperature dependence separately. In the appendix, we outline how to distinguish changes caused by the sample and changes due to artifacts.

\subsection{Depth dependence} \label{subsec:depth_dependence}
When varying the muon implantation energy at base temperature, all samples exhibit the same trend, a clear increase in muon field and muon depolarization rate for the lowest and highest muon energies, as shown in Figure (\ref{fig:all_samples_depth} a,c). These are the energies where a measurable fraction of muons is implanted at the surface and substrate interface, respectively.\\ 
In the case of the muon field, the increase is similar for surface and interface in all three samples. For the iron-containing samples, the muon field at 1 keV and 25 keV increases by about 110-140 $\mu$T compared to the center of the film. The pure Pd sample displays about half the field increase, 70 $\mu$T. The average field value in the center of the film (8 - 14 keV) varies per sample, $151.260 \pm 0.003$ mT for Pd(Fe), $151.374 \pm 0.003$ mT for pure Pd and $150.916 \pm 0.003$ mT for Pd(Fe)Au. The last sample was measured in a different cryostat, which led to a slightly lower external magnetic field.\\
Regarding the depolarization rate, the iron-containing samples exhibit higher values with respect to pure Pd at all implantation energies, which is expected as a consequence of GMM formation. In the center of the film (8 - 14 keV), the average depolarization rate is $0.068 \pm 0.003$ $\mu s^{-1}$ and $0.077 \pm 0.003$ $\mu s^{-1}$for Pd(Fe) without and with gold, respectively and only $0.010 \pm 0.002$ $\mu s^{-1}$ for pure Pd. As outlined in the appendix, data sets from all three samples exhibit the same measurement artifact, a pronounced depolarization rate increase at the surface due to reflected muons. Whereas the surface depolarization rate due to the artifact can be enhanced by up to almost 0.1 $\mu s^{-1}$, the sample-induced increase at the interface is only on the order of 0.01 $\mu s^{-1}$ compared to the center of the film. \\
Besides the depolarization rate, also the measured asymmetry (depicted in Figure (\ref{fig:all_samples_depth} b)) varies for surface and interface. It decreases by a percentage corresponding to the percentage of backscattered muons at the surface of all samples, see the appendix for more details. At the interface, the asymmetry slightly decreases for the samples without gold. The asymmetry of the Pd(Fe) with gold stays constant. The slight decrease  for the samples without gold can be attributed to a small fraction of the muons forming muonium in the silicon substrate \cite{Prokscha2007}. Therefore we can conclude that a measurable fraction of muons reaches the silicon substrate and subsequently the sample/substrate interface for implantation energies of 21 keV and higher for Pd(Fe). For pure Pd, the interface is possibly reached for 21 keV, certainly for 23 keV and 25 keV. It is important to note that for both samples, these are exactly the implantation energies where an increase in depolarization rate is observed, while the muon field is already enhanced at 18 and 20 keV. For the sample with gold, we cannot use the asymmetry as an indicator for muons implanting in the gold layer as there is no muonium formation in gold, but we can state that muons do not reach the silicon substrate at the highest implantation energy. Comparing the depolarization rate and muon field values at the interface of the Pd(Fe) sample with gold, the muon field is maybe enhanced at 18 keV, whereas both depolarization rate and muon field show a clear increase for 20 keV and higher energies.

\begin{figure}
\centering
\includegraphics[width=\columnwidth]{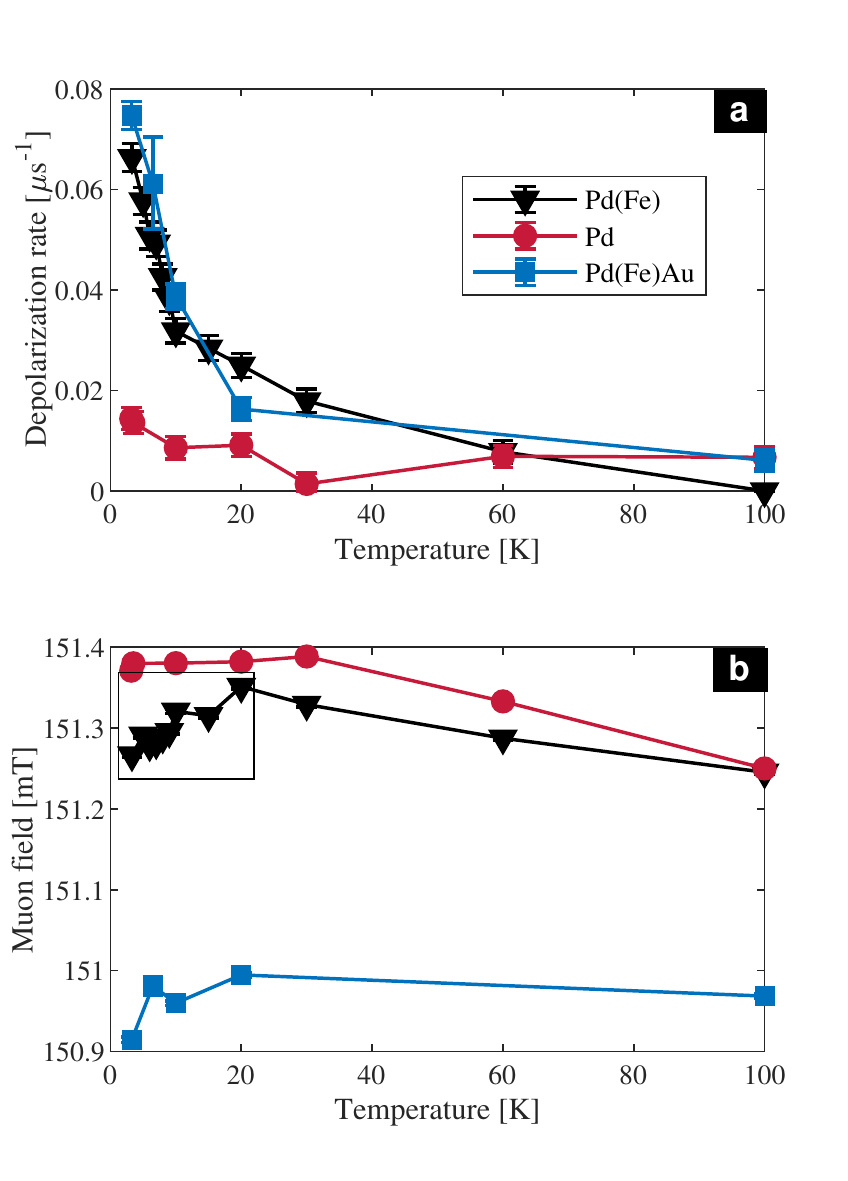}
\caption{Depolarization rate and muon field versus temperature for all samples at an implantation energy of 14 keV. Below 30 K, the increase in depolarization rate and the decrease in muon field are a clear sign of giant magnetic moments (GMM), as most clearly seen in the black box. The behavior above 30 K is dominated by an artifact caused by a slightly magnetic heating wire of the cryostat, see the appendix for a more detailed description of known measurement artifacts.}
\label{fig:all_samples_vs_temp}
\end{figure}

\begin{figure}
\centering
\includegraphics[width=\columnwidth]{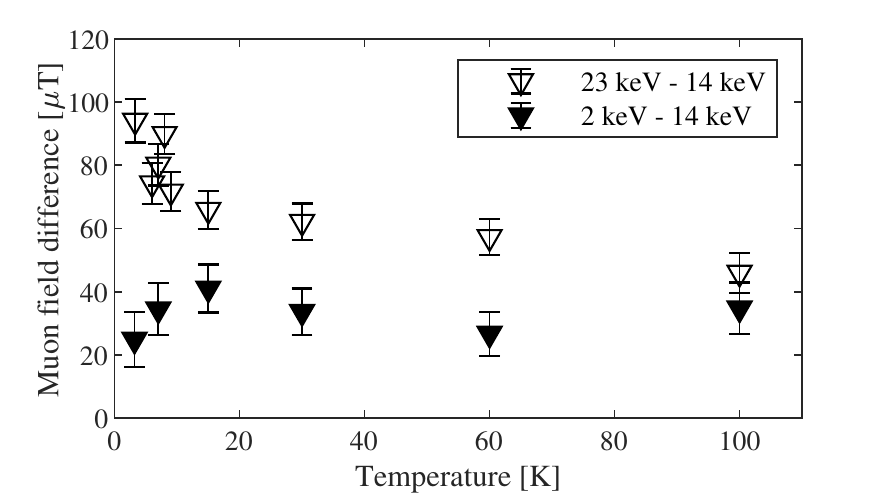}
\caption{Muon field increase at surface and interface versus temperature for sample 2 (Pd(Fe)). The muon field difference between 2 keV (considerable surface contribution) and 14 keV (bulk-like) as well as between 23 keV (considerable interface contribution) and 14 keV is nearly temperature-independent.}
\label{fig:difference_field_temp}
\end{figure}

\subsection{Temperature dependence}
The temperature dependence of the depolarization rate and the muon field in Figure (\ref{fig:all_samples_vs_temp}) shows apparent differences between the pure Pd samples and the two iron-containing samples. Lowering the temperature from 30 K to 3.7 K leads to an increase in depolarization rate from below 0.02 $\mu s^{-1}$ to around 0.07 $\mu s^{-1}$ for both Pd(Fe) samples. At the same time, the muon field reduces by 0.062 mT and 0.081 mT (sample without / with gold). The combined change in muon field and depolarization rate is a clear signature of GMM formation due to the iron impurities, as has been demonstrated by Nagamine et al. in TF-$\mu$SR measurements on bulk iron-doped Pd samples \cite{Nagamine1977a}.\\
The muon field of the pure Pd sample is temperature-independent, the depolarization rate increases only slightly upon lowering the temperature.\\
Figure (\ref{fig:difference_field_temp}) focusses on the temperature dependence of the muon field specifically at the surface and substrate interface. We use the bulk-like 14 keV values as reference points and subtract them from the values at 2 keV and 23 keV. It can be seen that, compared to the 14 keV values, the muon field at the surface (2 keV) and substrate interface (23 keV) is increased by tens of microteslas at all temperatures and that this increase is nearly temperature independent. Only the 23 keV data below 10 K shows a slight temperature-dependent trend, which could be ascribed to a difference in GMM contribution between the 14 keV and 23 keV data. Nevertheless, the data clearly shows the existence of a temperature-independent contribution to the muon field increase. Furthermore, we confirmed that the noticeably higher depolarization rate at the surface that we described in Section \ref{subsec:depth_dependence} is present for all temperatures (data not shown).

\section{Discussion}
The most prominent and unexpected feature of our data is the muon field increase at the surface and interface of all samples. This field increase shows five important characteristics that we can use to unravel the mechanism behind it. Namely, the field increase\\
\begin{itemize}
\item doesn't depend on temperature,
\item extends over several nanometers (see below),
\item is accompanied by an increase of field inhomogeneity,
\item is stronger in the iron-containing samples
\item and has a similar magnitude for Pd interfacing directly with silicon, with the gold spacing layer or with vacuum (at the surface).
\end{itemize}
In order to understand the origin of the muon field increase, we will first discuss how to determine the spatial extent of the effect. Secondly, we will take a closer look at the muon Knight shift in Pd. Thirdly, we will introduce potential mechanisms that can cause a magnetic field increase at transition metal surfaces / interfaces. In the last part of this section, we will compare our results to previous work using other measurement techniques on Pd thin films and nanoparticles.

\subsection{Spatial extent of the field increase}
The length scale of our measured field increase is an important criterion to rule out potential explanations. It is, however, not straightforward to determine this length scale, because most signals we measure are a superposition of two contributions. For example, for muons with implantation energies $\geq$18 keV, there is always one signal contribution from the center of the film and one from the interface region. Unfortunately, the signal to noise ratio does not allow to separate the two contributions in a two-component fit.\\
The fitted muon field, obtained with Equation (\ref{eq:poscountrate}), is therefore a weighted average of the field in the center and at the interface. It depends on the muon stopping profiles, the spatial extent of the interface region and the field that the muons experience there. A lower limit for the muon field at the interface  is given by the increase of the fitted field, which is between 110 and 140 $\mu$T for the Pd(Fe) films, and about 70 $\mu$T for the Pd film. We simulated the raw data created by the combination of two muon fractions: One with a field of 150 mT and one with a field of 150 mT $+ \Delta B$, where we varied the field increase $\Delta B$ between $200 \mu$T and $800 \mu$T. Then we performed a fit to the simulated data. These simulations reveal that for our experimental statistics, the muon fraction with the increased magnetic field needs to contain $\geq$5\% of all muons in order to produce a noticeable effect on the fitted muon field. The spatial extent of the region with enhanced local field can therefore be estimated from the stopping profiles calculated with TrimSP. Figure (\ref{figure:stopping}) shows the stopping profiles for sample 1. At the surface, the field shift appears for energies $\leq 3$ keV. This means that the width of the region with enhanced field must be about 3 nm: at 3 keV, the fraction of muons stopping in the top 3 nm is $< 5$ \%. The region at the interface appears much wider: the fitted muon field starts to increase between 18 and  21 keV, which means that the region of enhanced muon field extends from about 81 nm to 96 nm.\\ 
The estimated numbers suggest that  the surface region width of about 3 nm is much smaller compared to about 15 nm in the interface region. This is, however, not the case as our analysis so far did not take into account the surface and interface roughness. The surface region width is unaffected by roughness. Both surface and interface roughness, however, do have an effect on the measured interface region width. This is because the effective distance between interface and surface will vary within the average sample thickness plus / minus a contribution from the (surface + interface) roughness. So on average, the interface region appears to be smeared out, while locally, for a specific sample thickness, it is much thinner. To account for this, a sample thickness distribution has to be combined with the stopping profile when applying the 5\% criterion to determine the interface region width. We calculated the contribution of local sample thicknesses from the measured AFM surface roughness profile shown in Figure (\ref{figure:Sample_picture}). We can't precisely determine the interface roughness experimentally, therefore we assume it to be zero for an initial estimate. This leads to an upper limit for the local interface region width of 10.5 nm. Our SEM pictures clearly show a noticeable interface roughness, therefore the actual value for the local interface region width will likely be several nanometers smaller than 10.5 nm, coming close to the surface region width of 3 nm. 

\subsection{Field increase and muon Knight shift}
Discussing potential origins of the field increase is more easily done in terms of contributions to the muon Knight shift $K_{\mu}$. As explained in section \ref{chap:LEM-MuSR}, $K_{\mu}$ denotes the total shift of the measured magnetic field relative to a reference field, usually the external magnetic field. We do not know the exact value of the external magnetic field at the sample location, but as we are investigating a field difference between surface / interface and the center of the thin film, it suffices to know their  relative shifts with respect to each other. We therefore plot the muon Knight shift in Figure (\ref{fig:all_samples_vs_temp} d) relative to the magnetic field in the center of the film (instead of relative to the external magnetic field). This means that while the magnitude of our measured shift can be directly compared to muon Knight shift values from the literature, the sign can be different.\\ 
From literature on previous $\mu$SR experiments with pure and iron-doped Pd bulk samples it is known that muons implant at a Pd interstitial site \cite*{Nagamine1977,Nagamine1977a}. This leads to the following palladium muon Knight shift caused by various internal magnetic field contributions \cite*{Gygax1981,Schenck1981a, Nagamine1977,Nagamine1977a,Imazato1980}:
\begin{equation} \label{eq:K_mu_Pd}
K_{\mu} =K_{cont}^{s,dir}  + K_{cont}^{d,core}(T) + K_{orb} + K_{dia} + K_{dip} + K_{RKKY}(T).
\end{equation}
As indicated in Equation (\ref{eq:K_mu_Pd}), some of the Knight shift terms vary with temperature, others are temperature-independent. Furthermore, they have different signs. With respect to an external magnetic field, the direct shift from the contact hyperfine field of the nearly-free s-electrons $K_{cont}^{s,dir}$ is positive. The contact hyperfine contribution from the d-holes $K_{cont}^{d,core}(T)$ is mediated via core polarization and therefore negative. The heavy d-holes responsible for $K_{cont}^{d,core}(T)$ form a significant part of the Pd density of states \cite*{Rahman1978,Dye1981}. As the Fermi energy $E_F$ crosses a steep flank of a narrow d-band \cite*{Chen1989,Mueller1970}, the d-contribution to the density of states at the Fermi level is strongly temperature-dependent and so is the corresponding muon Knight shift term \cite{Gygax1981}. The orbital Knight shift $K_{orb}$ does not depend on the density of states, it is positive and temperature-independent \cite{Knightshift1976}. $K_{dia}$ is the diamagnetic Knight shift, it is negative. $K_{dip}$ comes from dipolar magnetic fields (from any magnetic moments in the vicinity), it can be positive or negative depending on the muon implantation site with respect to the magnetic moment. $K_{RKKY}$ is the shift caused by the spin polarization of the conduction carriers (mostly d-holes) due to impurity-induced RKKY interaction, in our case it is therefore only present in the iron-doped samples. $K_{RKKY}$ only becomes relevant at low temperatures when RKKY oscillations can form. For iron-doped Pd, $K_{RKKY}$ is negative and it starts to play a role well below 100 K \cite*{Nagamine1977, Nagamine1977a}.

\subsection{Potential mechanisms for surface / interface field increase}
Given the sign and temperature-independence of the observed field increase at the surface / interface, it could originate from orbital or spin moments that lead to an increased orbital and dipolar Knight shift. Another option are changes in the density of states that increase $K_{cont}^{s,direct}$  or reduce $K_{cont}^{d,core}(T)$ and $K_{RKKY}(T)$. In the following, we will discuss potential mechanisms in more detail. 

\subsubsection{Surface state with suppressed d-states}
Several authors \cite*{Shinohara2003,MacFarlane2013} have suggested the existence of an electronic surface state in Pd. Such a surface state could produce a temperature-independent field increase if it suppresses d-electron contributions ($K_{cont}^{d, core}$ and $K_{RKKY}$) as can be seen in Equation (\ref{eq:K_mu_Pd}).\\
A quantitative estimate, however, reveals that suppressed d-states cannot explain the magnitude of our observed field increase. For the case of pure Pd, we can refer to Gygax et al. \cite{Gygax1981}, who report a total muon Knight shift of around -300 ppm below 100 K, which they ascribe for a large part to $K_{cont}^{d, core}$. With an external field of 150 mT, 300 ppm would correspond to 45 $\mu$T. Our measured field increase clearly exceeds 45 $\mu$T for all samples and temperatures. One also has to consider that only a small fraction of the muons implants in the region where these 45 $\mu$T could be felt. The average measured field increase would therefore be much smaller than 45 $\mu$T, below the detection limit of the beamline. (On a side note: a similar argument holds for any potential contribution coming from $K_{cont}^{s,dir}$, as it is even smaller than $K_{cont}^{d, core}$.)\\
Furthermore, such surface states usually reside within the first few atomic layers only. A last argument against suppressed d-states is the expected effect on the field inhomogeneity: With suppressed d-states, we would expect a decrease in field inhomogeneity instead of an increase. Considering all the arguments against a surface state with suppressed d-states, we can conclude that this is not a good explanation for our observed signal.

\subsubsection{Adsorbents and Silicon Dangling Bonds}
We exposed the samples to air for an extended period of time, which unavoidably leads to the adsorption of molecules from the air, such as water and hydrocarbons. The adsorbed molecules possess spin magnetic moments, which leads to a $K_{dip}$ contribution with a negligible temperature dependence in our temperature range. (The temperature dependence is governed by the Boltzmann distribution.) These disordered magnetic moments should also lead to an increase in field inhomogeneity. Both statements are also true for unsaturated (dangling) bonds on the surface of the silicon substrate. In both cases it is important to notice that these magnetic moments are not located \textit{within} the sample, but \textit{on top} of it or directly \textit{below} it. This means that muons will be implanted at distances of up to 3 nm to those magnetic moments, which allows us to make an estimate for the expected signal magnitude based on the depth dependence of dipolar magnetic fields. Initially the estimates look promising:  At 1 nm distance from a localized electron spin with a magnetic moment of $\mu_B$, its dipolar magnetic field is about 1-2 mT (depending on where exactly the muon is located with respect to the spin). At 3 nm, this has dropped to less than 70 $\mu$T. One has to, however, consider that such a positive field increase will only be produced by electron magnetic moments aligned with the external magnetic field (spin down). An almost equal amount of magnetic moments is anti-aligned with the external field (spin up) and produces an equally big magnetic field decrease. The difference in numbers of up and down spins (the polarisation) is governed by the Boltzmann distribution. At 3.7 K, the polarization is only 1.4 \%, thereby reducing the expected field values by two orders of magnitude.\\
We can make a similar estimate for the $K_{dip}$ contribution of the iron impurities, with the same result: The expected muon field increase would be much smaller than the values we observe.\\
It might be possible that adsorbents and silicon dangling bonds polarize the Pd d-holes, similar to the giant magnetic moments that iron impurities form in Pd. This would increase the total magnetic moment size. It would, however, not necessarily lead to a net increase in Knight shift and therefore measured muon field, since other Knight shift contributions need to be taken into account as well.\\
In conclusion, adsorbents, silicon dangling bonds and iron-impurities at the surface / interface can in principle produce a positive, nearly temperature-independent magnetic field increase and an increase in field inhomogeneity. All dipolar field increases are, however, much too small in magnitude to explain our observed signal.

\subsubsection{Crystal irregularities}
Surface and interface of a thin film are prone to increased stress or strain and irregularities in crystal structure, which can result in additional localized magnetic moments and therefore an additional $K_{dip}$. In the case of Pd, one option for surface disorder is local hcp instead of fcc ordering, but also twinning, defects and stacking faults. Alexandre et al. performed DFT calculations and found an increase of the DOS at the Fermi level in the Pd hcp phase \cite{Alexandre2006}. Changes in the DOS primarily affect $K_{cont}^{d,core}(T)$. Furthermore, Alexandre et al. describe how local hcp ordering as well as other two-dimensional defects can introduce local magnetic moments, especially at surfaces.\\
As our films are polycrystalline with a rough surface, we can safely assume local perturbations of the fcc crystal structure at the surface. It is easily possible that these perturbations extend several nanometers into the film. Additional magnetic moments also increase local magnetic field inhomogeneities.  A field increase in our Pd samples due to crystal irregularities therefore seems plausible. On the other had, an effect related to a change in the DOS is likely to exhibit a temperature dependence, though it is difficult to predict its magnitude without extensive calculations. Within our measurement accuracy, we do not observe a temperature dependence. The proposed mechanism is furthermore intrinsic to Pd. There is no apparent reason why the samples with iron should develop more crystal irregularities, especially given the dilute iron concentration. One could imagine that the negative $K_{RKKY}$ is suppressed at the surface/interface, because RKKY oscillations cannot form so easily due to more defects in the crystal structure. We did, however, see clear indications of GMM formation and RKKY oscillations at the surface and interface of sample 2, namely the same features as shown in Figure (\ref{fig:all_samples_vs_temp}): a muon field decrease and a depolarization rate increase below 30 K.\\
In conclusion, local magnetic moments induced by crystal irregularities are a plausible explanation for our observed field increase, if we assume that their temperature dependance is smaller than our measurement accuracy. This scenario, however, offers no explanation for the additional field increase in the iron-doped samples. 
\subsubsection{Surface state with induced orbital moments via spin-orbit coupling}
A temperature-independent, positive Knight shift $K_{orb}$ is a signature of orbital moments \cite{Knightshift1976}. Orbital moments of conduction electrons at surfaces can be induced by various forms of spin-orbit coupling. In the bulk of transition metals, these are almost completely quenched because of symmetry arguments, but these arguments don't hold anymore at a surface \cite{Binns2001}.\\
Generally speaking, conduction electrons at the surface can get trapped in orbits determined by the surface properties. The influence of spin-orbit coupling is most easily understood in an energy-minimization picture: Analogous to the Zeeman effect, spin-orbit coupling leads to a splitting of conduction electron energy bands, with the two split bands corresponding to the spin and orbital moment being aligned and anti-aligned. As the state with the two moments aligned is energetically more favorable, it will be more populated, thereby inducing a net orbital moment. This effect is at the center of the research field of spin-orbitronics, which has grown rapidly during the last 10 years. A recent review of spin-orbitronics at transition metal interfaces is given by Manchon and Belabbes \cite{Manchon2017}.\\
One mechanism to induce orbital moments at a surface via spin-orbit coupling is the Rashba-Bychkov effect, observed originally in 2D semiconductors \cite{Manchon2015}. The presence of a surface breaks inversion symmetry, there is an electric surface potential. While moving in this electric potential, a conduction electron experiences a magnetic field in its rest frame, the spin of the electron couples to this magnetic field, thereby inducing an orbital moment. The Rashba-Bychkov effect has been described theoretically for metal surface states \cite{Bihlmayer2006} and observed experimentally for various heavy metals as they have strong spin-orbit coupling \cite{Wang2017}. Pd with a spin-orbit coupling constant of 187 meV \cite{Grytsyuk2016} is likely to exhibit a similar surface state. Furthermore, it has been demonstrated that the Rashba-Bychkov effect can be increased by impurity surface decoration, for a variety of surface atoms on heavy metal substrates \cite*{Rotenberg1999, Hochstrasser2002, Forster2004} as well as for residual CO and H$_2$O molecules \cite{King2011} and Fe atoms \cite{Wray2011} on a Bi$_2$Se$_3$ surface. The stronger field increase for our iron-doped samples therefore seems to fit a Rashba-Bychkov explanation, even though conclusions from other impurity / sample combinations should only carefully be transferred to Fe atoms on a Pd surface. A Rashba surface state, however, usually resides only within the first few atomic layers of a sample, therefore it cannot fully explain our observed field increase due to its larger spatial extent. (Our effect extends over about 3 nm at the surface and less than 10.5 nm at the interface.) \\
Another general mechanism to induce orbital moments at a surface is via localized charges / spins. The orbital motion of a conduction electron couples to the localized spin. In the case of a localized electric charge, the spin of the conduction electron couples to the magnetic field it experiences in its rest frame, because of moving in an electric potential, just as in the case of the Rashba-Bychkov effect. Of course, usually, both localized charge and spin are present. Orbital moments induced by localized spins and charges were found to be the origin of ferromagnetism in twinned Pd nanoparticles as well as thiol-capped gold nanoparticles and thin films \cite*{Hernando2006,Hernando2006a}. Induced magnetic moments in the thiol-capped gold films were gigantic, 10 - 100 $\mu_B$ per gold surface atom. Our Pd films certainly contain localized spins and charges in the surface / interface region as described in the paragraphs about crystal irregularities and surface adsorbents / silicon dangling bonds. We can therefore expect that orbital moments are induced in that region as well, on a length scale in accordance with our experimental observations for the field increase. Furthermore, iron impurities are localized spins / charges, so there should be more orbital moments induced in the iron-doped Pd samples, which also corresponds with our measurement results. We can conclude that all our experimental observations can be explained by induced orbital moments in the surface / interface region. More work is needed to confirm this conclusion by a quantitative estimate of the induced orbital moments and the resulting muon fields. 

\subsection{Comparison with previous work}
The most relevant experimental results for comparison are the ones by Parolin et al. \cite*{Parolin2006,Parolin2007,Parolin2009} and MacFarlane et al. \cite{MacFarlane2013}. Just as $\mu$SR, their $\beta$-NMR technique measures internal magnetic fields by implanting a local probe (a lithium ion) in the sample. They studied Pd thin films and foils with thicknesses ranging from 28 nm to 12.5 $\mu$m. Both in epitaxial and polycrystalline films, they observe a large negative, temperature- and thickness-dependent lithium Knight shift. Its temperature dependence follows the behavior of the magnetic susceptibility $\chi$. In the epitaxial films, they additionally observe a clearly visible  surface state that is temperature- and thickness-independent and both present in Au-capped and non-capped films. The same surface state is also observed in some of the polycrystalline samples, though less clearly visible due to broadened peaks in the resonance spectrum. The surface state is characterized by a small, constant, positive lithium Knight shift. They emphasize that the magnitude of the surface resonance is much too big to originate only from the topmost atomic layer. Furthermore, they name a likely explanation for this surface state: d-states are suppressed at the surface and only the s-states with a positive lithium Knight shift remain. Regarding other possible origins of their surface state, we can take similarities between polycrystalline and epitaxial samples as an argument to rule out a dominant role of crystal irregularities.\\
From the large amount of Pd-nanoparticle-related literature, we focus on descriptions of magnetic surface properties and on similarities and differences with thin film surface states.\\
Various authors report nanoparticle (NP) surfaces with temperature-independent magnetic susceptibility (paramagnetic NPs) or saturation magnetization (ferromagnetic NPs) \cite{Shinohara2003}. Hernando et al. suggested orbital moments of the surface conduction electrons to be the reason for a temperature-independent magnetic surface state in Pd nanoparticles \cite{Hernando2006}. The transition between surface and bulk-like properties in NP is often modeled in terms of a "healing length", an intermediate region where the material "heals" back to bulk-like properties. The surface is usually thought to consist of only a few monolayers, reported values for the width of the intermediate region (the "healing length") are usually below 1 nm \cite*{Volokitin1997,VanLeeuwen1992}. MacFarlane et al. \cite{MacFarlane2013} applied the same healing length model to their Pd thin film data, with the density of states healing from a reduced value at the surface to the normal bulk value. This resulted in a large surface thickness of 22.5nm and a healing length of 6 nm, much larger than the NP values. Their surface thickness is also much larger than our measured values of about 3 nm surface thickness and $<$10.5 nm interface thickness.\\
Our work confirms the existence of a temperature-independent magnetic surface state in Pd, a signature previously observed both in Pd nanoparticles and thin films. We gain further information about its spatial extent and on the depth-dependent magnitude of the local magnetic field.  We believe induced orbital moments to be the most likely origin of our surface state, in combination with induced magnetic moments from crystal irregularities. A surface state with orbital moments has so far only been considered for Pd NPs. We therefore hope that our observations can contribute to a more unified picture of Pd magnetic surface states, pointing towards an important role of orbital magnetic moments.

\section{Summary and outlook}
We investigated magnetic finite-size effects in three pure and iron-doped Pd thin films by means of systematic depth-dependent muon spin spectroscopy. From our $\mu$SR data we infer the presence of a magnetic surface state both at the film surface and at the interface with the substrate. This magnetic state extends over a few nanometers and it is characterized by a temperature-independent increase of the local magnetic field as well as a slight increase of local field inhomogeneity. We rule out suppressed d-states as an origin and demonstrate that magnetic moments from adsorbents can not produce a magnetic field large enough to explain our observations. Magnetic moments induced by crystal irregularities can qualitatively explain some of our findings. Orbital magnetic moments induced by localized spins and charges or by inversion symmetry breaking at the surface via de Rashba-Bychkov effect, however, can qualitatively explain all features of the observed surface state. We therefore consider it likely that we observed orbital surface moments with muon spin spectroscopy for the first time.\\
We hope that our work can help to better understand the magnetism of Pd nanoparticles used for catalysis and to engineer metal surface properties according to the needs of spin-orbitronics. Furthermore, it could be relevant in a multi-technique approach to reduce non-contact friction in the fields of magnetic resonance force microscopy and NV-center magnetometry, where magnetic moments at surfaces play an important role. Last but not least, we hope that our measurements can stimulate other systematic depth-dependent $\mu$SR studies of materials with extraordinary surface states, such as topological insulators. 

\section{Acknowledgments} 
The authors thank T. Benschop for help with sample fabrication, measurements and data analysis. We also acknowledge M. Hesselberth, D. Scholma and T. Mechielsen for technical support, N. Lebedev, A. Ben Hamida and J. Aarts for help with the SQUID measurements, M. Flokstra and A. Hernando for stimulating discussions and Z. Salman for support with MuSRfit. G.W. acknowledges financial support from Ev. Studienwerk Villigst. L.B. was supported by the Netherlands Organization for Scientific Research (NWO) through a VENI fellowship (0.16.Veni.188.040).

\section{Author Contributions}
G.W. and L.B. conceived the experiment. M.d.W. made the samples. G.W., M.d.W. and T.P. performed the measurements. A.S. contributed to the development and maintenance of the muon beamline. G.W. and T.P. did the analysis of the data. G.W. and J.M. wrote the manuscript. L.B. provided guidance in all stages of the project. All authors discussed the results and reviewed the manuscript.

\section*{Appendix}
\section{Distinguishing artifacts and signal} \label{app:artifact_signal}
Contributions of surface and interface effects to the measured $\mu$SR asymmetry spectra are expected to be small. Therefore, it is of utmost importance to understand measurement artifacts and to distinguish them from the actual sample signal.\\
One known artifact is a small offset to the external magnetic field on the order of 0.1 mT, which has been measured as well for other samples and by other users of the beamline. It stems from the sample environment, most likely a heating wire of the used LEM cryostat. This additional field decreases with increasing temperature above 30 K. We observed this characteristic temperature dependence in all temperature scans we performed, independent of sample and muon energy, as can be seen in Figure (\ref{fig:all_samples_vs_temp}). As the additional magnetic field is homogeneous on the length scale of the sample thickness, it doesn't have any influence on our comparison between surface, inner part and interface of the sample.\\
Another artifact comes from the varying degree of spin polarization of the muon beamline, depending on the transport settings of the muon beam. This effect is intrinsic to the beamline and well understood. In our experiment, we used two different beam transport settings for implantation energies above and below 8 keV. For the transport settings used for energies $>8$ keV, the beam polarization is  about 25\% higher. In Figure (\ref{fig:all_samples_depth} b), this artifact has been corrected by multiplying the measured asymmetries for the lower muon energies by the factor  1.25. This correction has no influence on the measured local field at the muon site and depolarization rate. The observed decrease of the asymmetry below 10 keV is due to the increasing probability for backscattering of the muons: the backscattered muons capture an electron and form the hydrogen-like muonium atom. In muonium the muon spin quickly depolarizes due to the hyperfine interaction with the electron, causing a reduction of the polarization of the muon ensemble, and therefore a reduction of observable decay asymmetry.\\
For 1, 2, and 3 keV, there is an additional artifact of reflected muons: due to the large decelerating electric field at these low implantation energies, some of the muons are being reflected before reaching the sample. The reflected muons stop in a large area of the radiation shield of the LEM cryostat. These muons experience the off-center field inhomogeneities of the LEM magnet, causing a "broader field distribution seen" by the muon ensemble, which in turn yields an increase of the observed depolarization rate. We measured the fraction of reflected muons with the help of a nickel plate mounted instead of the sample. Due to the ferromagnetic nature of nickel, muons reaching the sample will depolarize extremely quickly (approximately within 100 ns), such that the afterwards observed precession signal originates from reflected muons. These measurements confirm that at an external field of 150 mT, 40\% of the muons are reflected at 1 keV, 24\% at 2 keV and 15\% at 3 keV. For the 14 keV measurements, the fraction of backscattered muons is less than 4\%. The results of the nickel measurements are confirmed by a beam transport simulation of the LEM apparatus using the GEANT4 based musrSim package \cite*{Sedlak2012,Xiao2017}.  In the simulation, at 1 keV, the artifact leads to a shift of the measured "local field" of about 60 $\mu$T due to the 40\% fraction of the muons stopping in the radiation shield, where the field of the LEM magnet is slightly larger. The magnetic field inhomogeneities "seen" by the muons stopping in the radiation shield causes an increase of the observed depolarization rate of about 0.07 $\mu$s$^{-1}$. This means that the depolarization rate at the lowest implantation energies is dominated by the artifacts from reflected muons. For the magnetic field, the situation is more subtle. In the case of pure Pd, the measured field increase is approximately of the same magnitude as the predicted artifact contribution. This can mean two things: Either, the field increase is primarily caused by reflected muons or the field increase due to reflected muons and the sample are of the same order of magnitude, given that the total measured field is always a weighted average of all field contributions. For the two iron-doped samples, however, the measured field increase is almost two times the expected artifact contribution and we conclude that the field increase is dominated by the sample contribution. Assuming the same origin of the field increase for all samples, we consider it likely that also the pure Pd sample does indeed exhibit an increased magnetic field in the surface region. This assumption is further substantiated when looking at the field increase at the interface. The observed field increase is symmetric at the surface and interface for all samples, which points towards the same origin. As we are sure that there is no artifact present at the interface (no muons are backscattered or reflected for higher muon energies), it seems rather unlikely that a field increase of the same magnitude would be purely due to an artifact at the surface. In conclusion, the depolarization data at low muon energies doesn't provide information about the sample, while the field increase does. At high muon energies, the muon field and depolarization rate increase can unambiguously be linked to the sample. The increase is therefore an interface effect, independent of any artifact.

\bibliography{MuSR_PdFe_Bibliography}
\bibliographystyle{apsrev4-2}

\end{document}